\documentclass[11pt]{article}

\usepackage{amsmath,amsthm,amssymb,cite,enumerate,graphicx,epsfig} 

\usepackage[usenames]{color}
\usepackage{bm}
\usepackage{authblk} 

\def \BEA { \begin{eqnarray}}
\def \EEA {\end{eqnarray}}
\def \BE {\begin{equation}}
\def \EE {\end{equation}}
\def\d{\mathrm{d}}

\def \WDS #1 {\mbox{$\Phi_{#1}^{S}$}}
\def \WDA #1 {\mbox{$\Phi_{#1}^{A}$}}
\def \WD #1 {\mbox{$\Phi_{#1}$}}

\def\a{\alpha}

\def\e{\epsilon}

\def \mi {\stackrel{i}{m}}
\def \mj {\stackrel{j}{m}}
\def \mk {\stackrel{k}{m}}
\def \mr {\stackrel{r}{m}}
\def \ms {\stackrel{s}{m}}
\def \mz {\stackrel{z}{m}}
\def \mq {\stackrel{q}{m}}
\def \mo {\stackrel{o}{m}}
\def \mD {\stackrel{2}{m}}
\def \mT {\stackrel{3}{m}}
\def \mC {\stackrel{4}{m}}

\def \mio #1 {\mi_{#1}\ ^{  \! \! \! \! 0}} 
\def \mjo #1 {\mj_{#1}\ ^{  \! \! \! \! 0}} 
\def \mko #1 {\mk_{#1}\ ^{  \! \! \! \! 0}} 
\def \mro #1 {\mr_{#1}\ ^{  \! \! \! \! 0}} 
\def \mso #1 {\ms_{#1}\ ^{  \! \! \! \! 0}} 
\def \mpo #1 {\mp_{#1}\ ^{  \! \! \! \! 0}} 
\def \mzo #1 {\mz_{#1}\ ^{  \! \! \! \! 0}} 
\def \mqo #1 {\mq_{#1}\ ^{  \! \! \! \! 0}} 
\def \moo #1 {\mo_{#1}\ ^{  \! \! \! \! 0}} 
\def \mDo #1 {\mD_{#1}\ ^{  \! \! \! \! 0}} 
\def \mTo #1 {\mT_{#1}\ ^{  \! \! \! \! 0}} 
\def \mCo #1 {\mC_{#1}\ ^{  \! \! \! \! 0}} 

\def \miJ #1 {\mi_{#1}\ ^{  \! \! \! \! (1)}} 
\def \mjJ #1 {\mj_{#1}\ ^{  \! \! \! \! (1)}} 
\def \mkJ #1 {\mk_{#1}\ ^{  \! \! \! \! (1)}} 
\def \mrJ #1 {\mr_{#1}\ ^{  \! \! \! \! (1)}}

\def \hbm #1 {\mbox{\boldmath{$\hat m^{(#1)}$}}}

\newcommand{\be}{\begin{equation}}
\newcommand{\ee}{\end{equation}}
\newcommand{\beqn}{\begin{eqnarray}}
\newcommand{\eeqn}{\end{eqnarray}}
\newcommand{\pa}{\partial}
\newcommand{\ba}{\begin{array}}
\newcommand{\ea}{\end{array}}
\newcommand{\pp}{{\it pp\,}-}

\def \BEAH {\begin{eqnarray*}}
\def \EEAH {\end{eqnarray*}}
\def \BEA {\begin{eqnarray}}
\def \EEA {\end{eqnarray}}
\def \BDM {\begin{displaymath}}
\def \EDM {\end{displaymath}}

\def \BR {{Levi-Civita--Bertotti--Robinson }}
\newcommand{\AS}{Aichelburg-Sexl }

\begin{document}

\title{\bf Ultrarelativistic boost of a black hole in the magnetic universe of \BR}

\author[1,2]{Marcello Ortaggio\thanks{ortaggio(at)math(dot)cas(dot)cz}}

\affil[1]{Institute of Mathematics of the Czech Academy of Sciences, \newline \v Zitn\' a 25, 115 67 Prague 1, Czech Republic}
\affil[2]{Instituto de Ciencias F\'{\i}sicas y Matem\'aticas, Universidad Austral de Chile, \newline Edificio Emilio Pugin, cuarto piso, Campus Isla Teja, Valdivia, Chile}

\author[3]{Marco Astorino\thanks{Marco(dot)Astorino(at)Gmail.com}}
\affil[3]{Departamento de Ciencias, Facultad de Artes Liberales,
 UAI Physics Center, \newline Universidad Adolfo Iba\~nez, Av. Padre Hurtado 750, Vi\~na del Mar,
Chile}

\date{\today}

\maketitle

\begin{abstract}

We consider an exact Einstein-Maxwell solution constructed by Alekseev and Garcia which describes a Schwarzschild black hole immersed in the magnetic universe of Levi-Civita, Bertotti and Robinson (LCBR). After reviewing the basic properties of this spacetime, we study the ultrarelativistic limit in which the black hole is boosted to the speed of light, while sending its mass to zero. This results in a non-expanding impulsive wave traveling in the LCBR universe. The wave front is a 2-sphere carrying two null point particles at its poles -- a remnant of the structure of the original static spacetime.
It is also shown that the obtained line-element belongs to the Kundt class of spacetimes, and the relation with a known family of exact gravitational waves of finite duration propagating in the LCBR background is clarified. In the limit of a vanishing electromagnetic field, one point particle is pushed away to infinity and the single-particle \AS \pp wave propagating in Minkowski space is recovered.

\end{abstract}

\bigskip

\section{Introduction}

\label{sec_intro}

Gravitational fields generated by fast-moving sources have been of interest for several years, especially in the mathematical study of gravitational waves \cite{piranifast}. By boosting the Schwarzschild line-element to the speed of light, 
Aichelburg and Sexl were able to construct an exact solution describing the gravitational field of a massless point particle \cite{AicSex71}. In the limit, the spacetime curvature concentrates on a null hyperplane, and the resulting metric belongs to the class of impulsive \pp wave \cite{Penrose68twist}. The special properties of the \AS solution make it relevant also to, e.g., the study of of high-speed black hole encounters \cite{DEath78} and Planckian scattering \cite{tHooft87}. 

The method of \cite{AicSex71} has been employed by a number of authors to describe the field of various ultrarelativistic objects in different backgrounds, mostly asymptotically flat ones (cf., e.g., \cite{GriPodbook} and references therein). A particularly interesting extension consists in the inclusion of a cosmological constant, as obtained in \cite{HotTan93} by an ultrarelativistic boost of the Schwarzschild-(A)dS spacetime. In this case, the curvature of the (A)dS background also affects the geometry and global properties of the resulting impulsive wave (cf. also \cite{PodGri97}), which is still non-expanding but is not a \pp wave anymore. Similarly, one could expect that, e.g., also the presence of external fields 
might affect the ultrarelativistic limit of static sources. It is of some interest, in particular, the case of black holes under the influence of an electromagnetic field, described by spacetimes which are neither asymptotically flat nor (A)dS. 

An exact solution of the Einstein-Maxwell equations representing a Schwarzschild black hole immersed in a spatially homogeneous electromagnetic field was constructed by Alekseev and Garcia in \cite{AleGar96}. An interesting feature of this spacetime is that it asymptotes the \BR (LCBR) universe \cite{LeviCivita17BR,Bertotti59,Robinson59}. The LCBR spacetime is the direct product AdS$_2\times S_2$ of a two-dimensional anti-de~Sitter space with a two-dimensional sphere with the same radius. It admits a six-dimensional group of continuous isometries $SO(1,2)\times SO(3)$ (being, in particular, static,  spherically symmetric and homogenous) and is conformally flat. Despite its simplicity, the LCBR geometry has found various physical applications since, for example, it describes the near-horizon geometry of extremal Reissner-Nordstr\"{o}m black holes \cite{Carter73} as well as the interaction region in the collision of two shock electromagnetic plane waves  \cite{BelSze74} (cf. \cite{Stephanibook,GriPodbook} for further comments and references).

The main purpose of the present paper is to derive the gravitational field generated by the Schwarzschild-LCBR black hole of \cite{AleGar96} when it moves ``with the speed of light'' (as defined in the following), and to describe the corresponding geometry. As we will show, this consists of an impulsive gravitational wave propagating in the LCBR universe and possessing a spherical, non-expanding wave front. Another purpose of this work is thus to relate this spacetime to a more general class of exact impulsive waves which were constructed previously in \cite{Ortaggio02,OrtPod02} using a geometric approach, and hence to provide a more physical interpretation of those. The relation with the family of Kundt spacetimes and with previously found solutions describing gravitational waves of finite duration in the LCBR universe \cite{GarAlvar84} will be also clarified.

We begin in section~\ref{sec_AG} by presenting the solution of \cite{AleGar96} and reviewing its basic features. This section is largely a summary of \cite{AleGar96}, but we also add a few new observations (such as the asymptotic properties, the form of the line-element near the axis and near the curvature singularity, the behaviour of certain invariants near the latter, and a comment on the Petrov type of the spacetime).
In section~\ref{sec_boost} we perform the ultrarelativistic boost of the line-element of \cite{AleGar96}, which leads to an impulsive wave propagating in the \BR universe. The geometry of the latter and its relation with previosuly investigated families of exact gravitational waves is discussed in the final section~\ref{sec_discussion}.

\section{The Alekseev-Garcia solution}

\label{sec_AG}

\subsection{The line-element}

The Alekseev-Garcia spacetime was obtained using a generating technique based on the monodromy data transform \cite{AleGar96}. The line-element reads
\be
 \d s^2=-e^{2\psi}\cosh^2\frac{z}{b}\d t^2+e^{2\gamma}(\d z^2+\d\rho^2)+e^{-2\psi}b^2\sin^2\frac{\rho}{b}\d\phi^2 ,
 \label{AG}
\ee
with
\beqn
 & & e^{2\psi}=\frac{\left(R_+ +R_- -2m\cos\frac{\rho}{b}\right)^2}{(R_+ + R_-)^2-4m^2} , \qquad e^{2\gamma}=\frac{\left(R_+ +R_- -2m\cos\frac{\rho}{b}\right)^2}{4R_+ R_-}\left[\frac{R_+ -b\sinh\frac{z}{b}+(l+m)\cos\frac{\rho}{b}}{R_- -b\sinh\frac{z}{b}+(l-m)\cos\frac{\rho}{b}}\right]^2 , \nonumber \\
 & & R_{\pm}^2=\left(l\pm m-b\sinh\frac{z}{b}\cos\frac{\rho}{b}\right)^2+b^2\cosh^2\frac{z}{b}\sin^2\frac{\rho}{b} ,
 \label{coeffs}
\eeqn
where $m,b,l$ are constant parameters. Together with the electromagnetic field $F_{\mu\nu}=A_{\nu,\mu}-A_{\mu,\nu}$ defined by the potential $A_\mu\d x^\mu=A_\phi\d\phi$  (up to an arbitrary constant duality rotation), where 
\be
 A_\phi=-b\frac{R_+ +R_- +2m}{R_+ +R_- -2m\cos\frac{\rho}{b}}\left(1-\cos\frac{\rho}{b}\right) ,
 \label{A}
\ee 
metric~\eqref{AG} is a solution of the Einstein-Maxwell equations. It clearly admits the two commuting Killing vector fields $\pa_t$ and $\pa_\phi$. 
Computing the curvature tensor leads to very complicated expressions. Nevertheless, computer-aided computation shows that the invariant condition $I^3-27J^2=0$ \cite{Stephanibook} is satisfied at various generic points and for various values of the parameters, indicating (thanks to its staticity) that the spacetime is generically of Petrov type~D (it may be more special at special points). 
The physical meaning of the three parameters can be understood by considering various limiting cases or the neighborhood of certain points, as we now discuss. Without loss of generality, hereafter we shall assume $b,m>0$.\footnote{The metric is manifestly invariant under $b\mapsto -b$, which only changes the sign of $A_\phi$. A change in the sign of $m$ can be compensated by redefining $l\mapsto -l$ and $\rho\mapsto \pi-\rho$, which again leaves the metric invariant and changes the sign of the potential (up to a gauge term).\label{footn_constants}}

\subsection{Asymptotics}

\label{subsec_asympt}

In the limit $z\to-\infty$, one has $R_+\sim R_-\sim b\cosh{\frac{z}{b}}$, so that both $e^{2\psi}\to 1$ and $e^{2\gamma}\to 1$. Therefore, the line-element~\eqref{AG} tends asymptotically to the LCBR AdS$_2\times S_2$ metric
 \be
 \d s^2_0=-\cosh^2\frac{z}{b}\d t^2+\d z^2+\d\rho^2+b^2\sin^2\frac{\rho}{b}\d\phi^2 ,
 \label{BR}
\ee
while $A_\phi\to A^0_\phi=-b\left(1-\cos\frac{\rho}{b}\right)$. This means $F_{\mu\nu}\d x^\mu\d x^\nu=-\sin\frac{\rho}{b}\d\rho\wedge\d\phi$, so that the constant $1/b^2$ parametrizes the the field strength $F_{\mu\nu}F^{\mu\nu}=2/b^2$. In order to have a regular asymptotic behaviour, it is thus natural to define the range of the angular coordinates $(\rho,\phi)$ as
\be
 \rho\in[0,\pi b] , \qquad \phi\in[0,2\pi] .
 \label{range}
\ee

However, in the opposite spatial direction $z\to+\infty$ one obtains instead asymptotically
 \be
 \d \hat s^2_0=-\cosh^2\frac{z}{b}\d t^2+\alpha^2(\d z^2+\d\rho^2)+b^2\sin^2\frac{\rho}{b}\d\phi^2 ,
 \label{BR2}
\ee
with
\be
 \alpha=\frac{(l+m)^2+b^2}{(l-m)^2+b^2} .
 \label{alpha}
\ee
With the choice~\eqref{range}, this asymptotic region thus describes a LCBR spacetime with a conical singularity at the poles $\rho=0,\pi b$, parametrized by $\alpha$ -- the conical singularity is absent if either $l=0$ or $m=0$. Locally, the geometry \eqref{BR2} is equivalent to \eqref{BR} up to rescaling $b\mapsto \a b$ in the latter (for this reason, $F_{\mu\nu}F^{\mu\nu}=2/(\a b)^2$ in the asymptotic region \eqref{BR2}).

\subsection{LCBR and Schwarzschild limits}

\label{subsec_limits}

For the special choice $m=0$, solution~\eqref{AG} reduces everywhere to the LCBR universe \eqref{BR}. On the other hand, in the limit of a vanishing Maxwell field (i.e., $b\to+\infty$), from \eqref{AG} one obtains the (exterior) Schwarzschild line-element in Weyl's coordinates \cite{Stephanibook,GriPodbook}, i.e.,
\be
 \d s^2=-e^{2\psi}\d t^2+e^{2\gamma}(\d\tilde z^2+\d\rho^2)+e^{-2\psi}\rho^2\d\phi^2 ,
 \label{S_W}
\ee
with
\be
 e^{2\psi}=\frac{R_+ +R_- -2m}{R_+ +R_- +2m} , \qquad e^{2\gamma}=\frac{\left(R_+ +R_- +2m\right)^2}{4R_+ R_-} , \qquad R_{\pm}^2=\left(\pm m-\tilde z\right)^2+\rho^2 .
\ee
Here $m$ is the standard Schwarzschild mass. Note that we have rescaled $\tilde z=z-l$ in order to get rid of the (now fictitious) parameter $l$. However, in order to describe the entire exterior region, one needs $\rho\in(0,+\infty)$ (in contrast to \eqref{range}), with $\rho=0$, $-m<\tilde z<m$ corresponding to the horizon (the well-known ``rod''). Recall that the standard Schwarzschild coordinates $(r,\theta)$ are defined by $\tilde z=(r-m)\cos\theta$, $\rho^2=r(r-2m)\sin^2\theta$ (cf., e.g., \cite{AleGar96,GriPodbook}).

\subsection{Horizon and singularities}

\label{subsec_horizon}

Killing horizons are defined by possible points where $e^{2\psi}=0$ in \eqref{coeffs}. As it turns out \cite{AleGar96}, this occurs if, and only if,
\be
 \rho=0 , \qquad l-m\le b\sinh\frac{z}{b}\le l+m .
 \label{horizon}
\ee
In this range of $z$, for $\rho\approx0$, metric~\eqref{AG} at the leading order takes the form\footnote{This approximate ``near-horizon'' metric was already given in section~VII of \cite{AleGar96}, up to a typo (an extra factor $\rho^2$) in the function $f$ used there. It should be emphasized that the line-element at the limiting values of the interval \eqref{horizon} must be computed separately enforcing $b\sinh\frac{z}{b}=l\pm m$ {\em before} taking $\rho\approx0$ (as \eqref{nearH} would become singular). This can be done straightforwardly and we omit it for brevity, just noticing that a conical singularity appears at $b\sinh\frac{z}{b}=l+m$ (cf. related comments in the case $b\sinh\frac{z}{b}>l+m$ discussed in the following). An analogous comment applies to metric~\eqref{nearSing} and we will not repeat it there. Note that similar care is also needed if one wants to study the Schwarzschild metric \eqref{S_W} at $\tilde z=\pm m$ for $\rho\approx0$.\label{footn_horizon}}
\beqn
 & & \d s^2\approx \Sigma^2\left[-\frac{\rho^2}{4b^2}\d t^2+\frac{4b^2m^2}{\left[b^2+(l-m)^2\right]^2}(\d z^2+\d\rho^2)\right]+4b^2\Sigma^{-2}\cosh^2\frac{z}{b}\d\phi^2 , \label{nearH} \nonumber \\ 
 & & \Sigma^2=\frac{1}{b^2}\frac{\left(b^2-l^2+m^2+2lb\sinh\frac{z}{b}\right)^2}{m^2-\left(l-b\sinh\frac{z}{b}\right)^2} .
\eeqn
It follows that spatial sections of the horizon do not have constant Gaussian curvature. Nevertheless, the latter can be straightforwardly integrated over the horizon, which for $l=0$ gives $4\pi$, and thus the Euler characteristic is 2, in agreement with the spherical topology (via the Gauss-Bonnet theorem; for $l\neq0$ the conical singularity at $b\sinh\frac{z}{b}=l+m$ should be taken into account -- cf. footnote~\ref{footn_horizon} and the following comments). 
The horizon area can be computed easily and reads\footnote{Beware that this expression differs from the corresponding one obtained (for $l=0$) in section~VII of \cite{AleGar96}.}
\be
 {\cal A_H}=16\pi m^2\frac{b^2}{b^2+(l-m)^2} .
\ee

For values of $z$ outside the interval \eqref{horizon}, the condition $\rho=0$ describes an axis. The local metric for $\rho\approx0$ reads
\beqn
 & & \d s^2\approx -\Psi^{-2}\cosh^2\frac{z}{b}\d t^2+\Psi^{2}\left[\hat\alpha^2(\d z^2+\d\rho^2)+\rho^2\d\phi^2\right] , \label{rho=0} \nonumber \\ 
 & & \Psi^{2}=\frac{b\sinh\frac{z}{b}-l-m}{b\sinh\frac{z}{b}-l+m} , \qquad \hat\alpha=1 , \qquad \mbox{ for } b\sinh\frac{z}{b}<l-m , \label{axis} \\
 & & \Psi^{2}=\frac{b\sinh\frac{z}{b}-l+m}{b\sinh\frac{z}{b}-l-m} , \qquad \hat\alpha=\alpha , \qquad \mbox{ for } b\sinh\frac{z}{b}>l+m , \nonumber
\eeqn
with \eqref{alpha}. In the range $b\sinh\frac{z}{b}>l+m$, for $l\neq 0$ there is therefore a conical singularity (as noticed in \cite{AleGar96}) which extends all the way to $z\to +\infty$, in agreement with the comment following the asymptotic metric \eqref{BR2}.

At first sight, it may seem surprising that there is not a second horizon at the antipodal point $\rho=\pi b$. However, this can be understood in light of footnote~\ref{footn_constants} -- a positive mass $m$ at $\rho=0$ is mirrored by a negative mass $-m$ at $\rho=\pi b$ and one could thus expect a naked singularity there (see also the comments in \cite{AleGar96}). More precisely, one finds that there is a timelike curvature singularity at\footnote{This was found in section~VIII of \cite{AleGar96}, up to a typo (replace there $l\mapsto -l$).}
\be
 \tilde\rho\equiv \rho-\pi b=0 , \qquad -l-m\le b\sinh\frac{z}{b}\le -l+m ,
 \label{sing}
\ee
and the local metric for $\tilde\rho\approx 0$ takes the form
\beqn
 & & \d s^2\approx -\frac{\Pi^2}{\tilde\rho^2}\d t^2+\frac{16m^2\left[(l+m)^2+b^2\right]^2}{b^4 \Pi^{6}}\tilde\rho^4(\d z^2+\d\tilde\rho^2)+\Pi^{-2}\cosh^2\frac{z}{b}\tilde\rho^4\d\phi^2 , \label{nearSing} \nonumber \\ 
 & & \Pi^2=4\left[m^2-\left(l+b\sinh\frac{z}{b}\right)^2\right] .
\eeqn
One can verify that near the singularity \eqref{sing}, for instance, the following invariant diverges
\be
 F_{\mu\nu}F^{\mu\nu}=\frac{8b^2\left[m^2-\left(l+b\sinh\frac{z}{b}\right)^2\right]^2}{m^2\left[(l+m)^2+b^2\right]^2}\frac{\cosh^2\frac{z}{b}}{\tilde\rho^2}+O(\tilde\rho^0) .
\ee
Computing curvature invariants is much more complicated but one can show, for example, that also $R_{\mu\nu}R^{\mu\nu}$ diverges at \eqref{sing}.

However, for values of $z$ outside the interval \eqref{sing}, $\rho=\pi b$ is not a curvature singularity but an axis. The local metric for $\tilde\rho\approx 0$ reads
\beqn
 & & \d s^2\approx -\Xi^{-2}\cosh^2\frac{z}{b}\d t^2+\Xi^{2}\left[\hat\alpha^2(\d z^2+\d\tilde\rho^2)+\tilde\rho^2\d\phi^2\right] , \label{rho=pib} \nonumber \\ 
 & & \Xi^{2}=\frac{b\sinh\frac{z}{b}+l+m}{b\sinh\frac{z}{b}+l-m} , \qquad \hat\alpha=1 , \qquad \mbox{ for } b\sinh\frac{z}{b}<-l-m , \\
 & & \Xi^{2}=\frac{b\sinh\frac{z}{b}+l-m}{b\sinh\frac{z}{b}+l+m} , \qquad \hat\alpha=\alpha , \qquad \mbox{ for } b\sinh\frac{z}{b}>-l+m , \nonumber
\eeqn
with \eqref{alpha}. Similarly as above, in the range $b\sinh\frac{z}{b}>-l+m$ there is a conical singularity if $l\neq 0$.

\section{Ultrarelativistic boost}

\label{sec_boost}

Before starting, let us observe that the isometries of the asymptotic LCBR spacetime include a natural notion of boost (defined by the AdS$_2$ factor, cf.~\eqref{boost} below). This makes it plausible that the \AS method can be adapted to the present context. We will show that this is indeed the case, also pointing out some technical differences.

First of all, it is useful to decompose the line element~(\ref{AG}) as
\be
 \d s^2=\d s_0^2+\Delta ,
 \label{BRdecomposition}
\ee
in which $\d s_0^2$ is the LCBR metric \eqref{BR}, and 
\beqn
 \Delta=4m\left[x\left(R_+ +R_- -mx\right)-m\right]\left[\frac{(1+\eta^2)}{(R_+ + R_-)^2-4m^2}\d t^2+\frac{b^2(1-x^2)}{\left(R_+ +R_- -2mx\right)^2}\d\phi^2 \right] \nonumber \\
 {}+\Bigg[\left(\frac{R_+ -b\eta +(l+m)x}{R_- -b\eta +(l-m)x}\right)^2\frac{(R_+ -R_-)^2+4mx(mx-R_+ -R_-)}{4R_+ R_-}\nonumber \\
 {}+\frac{\left(R_+ +R_- -2b\eta  +2lx\right)\left(R_+ -R_- +2mx\right)}{\left[R_- -b\eta +(l-m)x\right]^2} \Bigg] 
	(\d z^2+\d\rho^2) ,
 \label{BRperturbation}
\eeqn
where we have introduced the compact notation
\be
 \eta=\sinh\frac{z}{b} , \qquad x=\cos\frac{\rho}{b} .
 \label{eta_x}
\ee

A notion of boost is defined w.r.t. the AdS$_2$ factor of the LCBR ``background'' $\d s^2_0$, i.e.,
\be
 \d s^2_{AdS_2}=-\cosh^2\frac{z}{b}\d t^2+\d z^2 ,
 \label{AdS2}
\ee
and leaves $\d s^2_0$ invariant. Therefore, we need only study how the term $\Delta$ in \eqref{BRdecomposition} transforms. To simplify later computations it is also useful to employ \eqref{AdS2} to write $\cosh^2\frac{z}{b}\d t^2=-\d s^2_{AdS_2}+\d z^2$, so that in \eqref{BRperturbation} one can substitute
\be
 (1+\eta^2)\d t^2=-\d s^2_{AdS_2}+\d z^2 .
 \label{dt2}
\ee 
(This has the advantage that no terms proportional to $\d t^2$ will appear in $\Delta$, except for the one contained in the boost-invariant quantity $\d s^2_{AdS_2}$.)
It is also convenient to embed AdS$_2$ in a flat 3-dimensional space, so that \eqref{AdS2} becomes
\be
 \d s^2_{AdS_2}=-\d {Z_0}^2+\d {Z_1}^2-\d{Z_2}^2 , \qquad -{Z_0}^2+{Z_1}^2-{Z_2}^2=-b^2 \ ,
 \label{AdS2_emb}
\ee
where 
\be
  Z_0=b\cosh\frac{z}{b}\cos\frac{t}{b} \ , \qquad Z_1=b\sinh\frac{z}{b} \ , \qquad Z_2=b\cosh\frac{z}{b}\sin\frac{t}{b} .
	\label{Zi}
\ee
In double null coordinates 
\be
 Z_0=\frac{U-V}{\sqrt{2}} , \qquad Z_1=\frac{U+V}{\sqrt{2}} ,
 \label{UV}
\ee
eq.~\eqref{AdS2_emb} becomes
\be
 \d s^2_{AdS_2}=2\d U\d V-\d{Z_2}^2 , \qquad -2UV+{Z_2}^2=b^2 .
 \label{AdS2_UV}
\ee
Using \eqref{Zi} and \eqref{UV}, we can replace $\d z^2$ in \eqref{BRperturbation} and \eqref{dt2} by
\be
 \d z^2=\frac{b^2}{2}\left[b^2+\frac{(U+V)^2}{2}\right]^{-1}(\d U+\d V)^2 .
 \label{dz2}
\ee

We are thus finally ready to perform the AdS$_2$ boost (i.e., a Lorentz boost in the direction $Z_1$), which in the coordinates \eqref{AdS2_UV} takes the simple form
\be
 U\mapsto\epsilon^{-1} U , \qquad V\mapsto\epsilon V ,
 \label{boost}
\ee
where $\epsilon>0$ (the same boost in the coordinates $(t,z)$ was discussed in \cite{Alekseev17}). An {\em ultrarelativistic boost} consists in taking the limit to the speed of light $\e\to0$, while simultaneously rescaling the mass as
\be
 m\mapsto\e p ,
\ee
such that the total energy remains finite \cite{AicSex71}.

The term \eqref{BRperturbation}, now parametrized by $\e$, thus becomes
\beqn
 \Delta_\e=4\e p\left[x\left(R^\e_+ +R^\e_- -\e px\right)-\e p\right]\left[\frac{-\d s^2_{AdS_2}+\frac{b^2}{2}(\e^{-1}\d U+\e\d V)^2(b^2+y_\e^2)^{-1}}{(R^\e_+ + R^\e_-)^2-4\e^2 p^2}+\frac{b^2(1-x^2)}{\left(R^\e_+ +R^\e_- -2\e px\right)^2}\d\phi^2 \right] \nonumber \\
 {}+\left[\left(\frac{R^\e_+ -y_\e  +(l+\e p)x}{R^\e_- -y_\e  +(l-\e p)x}\right)^2\frac{(R^\e_+ -R^\e_-)^2+4\e px(\e px-R^\e_+ -R^\e_-)}{4R^\e_+ R^\e_-}\right. \nonumber \\
  {}+\left.\frac{\left(R^\e_+ +R^\e_- -2y_\e   +2lx\right)\left(R^\e_+ -R^\e_- +2\e px\right)}{\left[R^\e_- -y_\e  +(l-\e p)x\right]^2} \right] 
	\left[\frac{b^2}{2}(\e^{-1}\d U+\e\d V)^2(b^2+y_\e^2)^{-1}+\d\rho^2\right] ,
 \label{BRperturbation_2}
\eeqn
where 
\be
 y_\e=\frac{\e^{-1}U+\e V}{\sqrt{2}} , \qquad (R^\e_{\pm})^2=\left(l\pm \e p-y_\e x\right)^2+(b^2+y_\e^2)(1-x^2) .
\ee
(Note that the shortcut $y_\e$ fully determines how ${\Delta_\e}$ depends on $U$ -- this will be important in subsequent calculations.)

We are now ready to compute the limiting metric ${\d s^2=\d s_0^2+\lim_{\e\to 0}\Delta_\e}$. By inspection, one can see that in the limit $\e\to 0$ only (some of) the terms proportional to $\d U^2$ will survive. It is also useful to note that
\be
  R^\e_{\pm}=R_\e\pm \e p\frac{l-y_\e x}{R_\e}+\ldots , \qquad R_\e\equiv\sqrt{\left(l-y_\e x\right)^2+(b^2+y_\e^2)(1-x^2)} .
\ee
where the dots denote terms which become negligible in the limit. For a small $\e$ one thus has 
\beqn
 \Delta_\e=2pb^2\d U^2\e^{-1}\frac{1}{(b^2+y_\e^2)(R_\e -y_\e  +lx)}\left(\frac{l-y_\e x}{R_\e}+x \right) +\ldots .
 \label{BRperturbation_3}
\eeqn

Applying the distributional identity
\be
 \lim_{\e\to 0} \frac{1}{\e}f\left(y_\e \right)
  =\sqrt{2}\,\delta(U)\int_{-\infty}^{+\infty}f(y)\d y ,
 \label{identity}
\ee
one arrives at the final metric  
\be
 \d s^2=\d s^2_0+2H\delta(U)\d U^2 , \qquad H=\sqrt{2}pb^2\int_{-\infty}^{+\infty}f(y)\d y ,
 \label{final}
\ee 
where (recall \eqref{AdS2_UV})
\be
 \d s^2_0=2\d U\d V-\d{Z_2}^2+\d\rho^2+b^2\sin^2\frac{\rho}{b}\d\phi^2 , \qquad -2UV+{Z_2}^2=b^2 ,  
 \label{backgr}
\ee 
and 
\be
 f(y)=\frac{1}{b^2+l^2}\frac{1}{b^2+y^2}\left(l+\frac{yl+b^2x}{\sqrt{y^2-2lxy+l^2+b^2(1-x^2)}}\right) .
 \label{f(y)}
\ee
with $x=\cos\frac{\rho}{b}$. The limit of the potential~\eqref{A} simply gives the potential of the LCBR solution (cf. section~\ref{subsec_asympt}), i.e.,
\be
 A_\phi=-b\left(1-\cos\frac{\rho}{b}\right) .
 \label{A_final}
\ee

It only remains to compute the integral in \eqref{final}. The first term in \eqref{f(y)} can be integrated elementarily. Using Euler's substitution $\sqrt{y^2-2lxy+l^2+b^2(1-x^2)}=\tau+y$ \cite{Gradshteynbook8} for the second term, one arrives at
\be
 \int f(y)\d y=\frac{1}{b^2+l^2}\left[\frac{l}{b}\arctan\frac{y}{b}+\frac{1}{2}\ln\frac{b^2(1+x)^2+\left(\sqrt{y^2-2lxy+l^2+b^2(1-x^2)}-y-l\right)^2}{b^2(1-x)^2+\left(\sqrt{y^2-2lxy+l^2+b^2(1-x^2)}-y+l\right)^2}\right] ,
\ee
so that, finally,
\be
 \d s^2=\d s^2_0+2H\delta(U)\d U^2 , \qquad H=\sqrt{2}p\frac{b^2}{b^2+l^2}\left(\frac{\pi l}{b}+\ln\frac{1+\cos\frac{\rho}{b}}{1-\cos\frac{\rho}{b}}\right) ,
 \label{final2}
\ee 
with \eqref{backgr}.

It is worth observing that, in the above calculation, the infinite gauge subtraction used in \cite{AicSex71} was not necessary. This is due to the contribution of the electromagnetic field, which makes the integral of \eqref{f(y)} finite.

\begin{figure}[t]
 \centering
 \includegraphics[scale=.3]{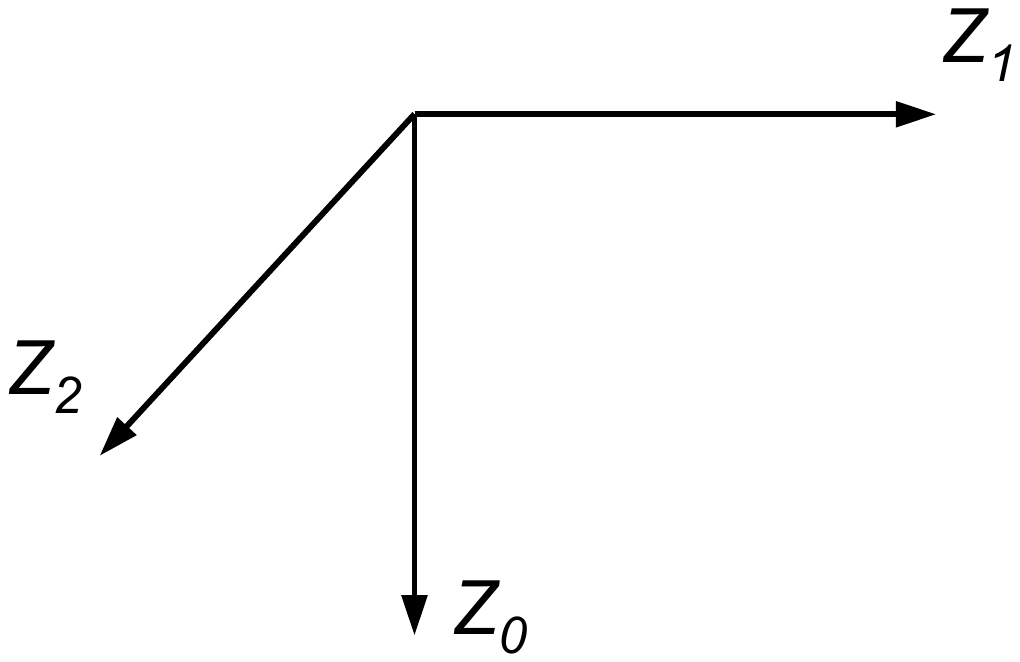}
\includegraphics[scale=.6]{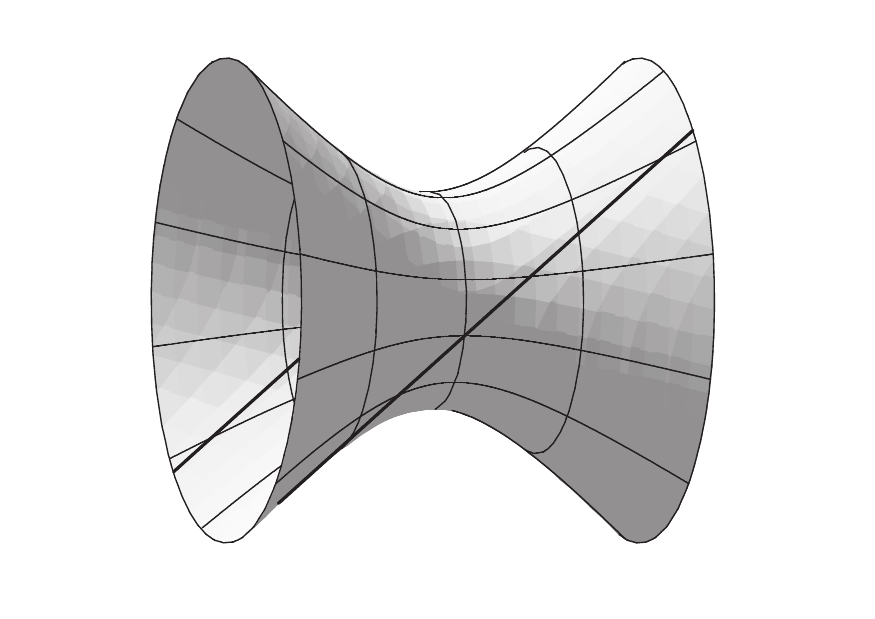}
 \caption{The 2-hyperboloid visualizes the dS$_2$ factor AdS$_2$ factor of the LCBR universe (\ref{BR}) in the embedding coordinates~\eqref{AdS2_emb}. Each point corresponds to a 2-sphere of a constant radius $b$ in the four-dimensional spacetime. The parallel solid lines $Z_0+Z_1=0$ ($\Leftrightarrow Z_2=\pm b$) are the histories of two of these spheres, propagating at the speed of light from one side of the universe to the other. Cf. \cite{PodGri97} for a similar discussion in the case of the solutions of \cite{HotTan93}.}
 \label{fig_hyperboloid}
\end{figure}

\begin{figure}[t]
 \centering
 \includegraphics[width=5cm]{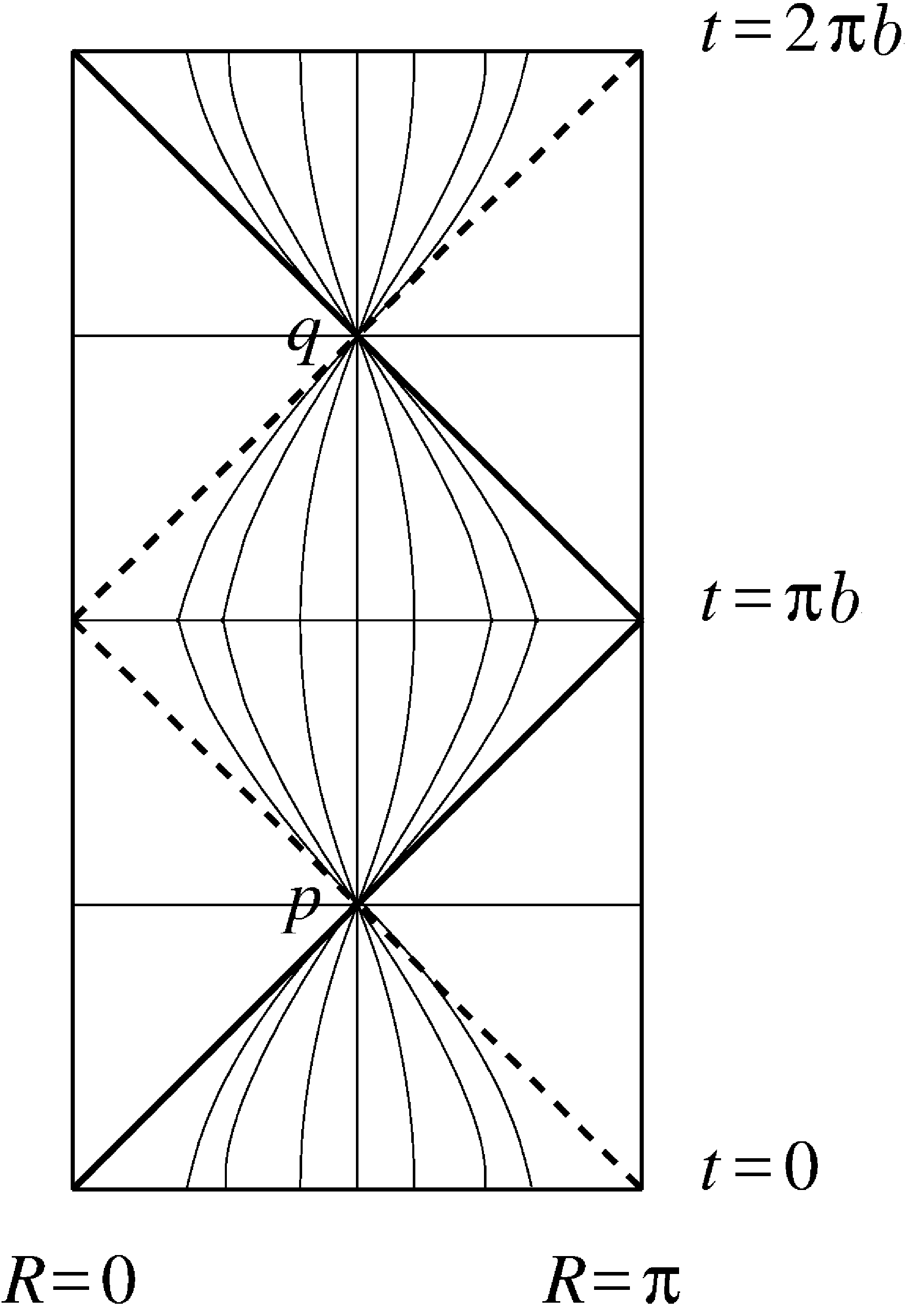}
 \caption{The conformal diagram of the LCBR universe (\ref{BR}) is that of its AdS$_2$ factor (cf. \cite{Carter73}), with each point 
	representing a two-dimensional sphere. The conformal spatial coordinate $R\in(0,\pi)$ is defined by $R=2\arctan(e^{{z}/b})$, and the timelike boundaries $R=0$ ($z=-\infty$) and $R=\pi$ ($z=+\infty$) correspond to null and spacelike infinity on opposite sides of the universe. The two solid lines represent the components $Z_2=\pm b$ of the null hypersurface $U=0$.  Timelike geodesics emanating from $p$ reconverge at the image point $q$ (cf. \cite{HawEll73}). 
	}
 \label{fig_bertotticonformal}
\end{figure}

\section{Discussion}

\label{sec_discussion}

The above spacetime \eqref{final2} with the magnetic potential \eqref{A_final} is an exact solution of the Einstein-Maxwell field equations. For $U\neq0$, it coincides with the LCBR solution~\eqref{BR} (in the embedding coordinates \eqref{backgr}). In addition, it contains an impulsive gravitational wave localized on the null 3-submanifold  $U=0$, and thus propagating in the LCBR universe at the speed of light (in the $z$-direction, recall \eqref{Zi}, \eqref{UV}). The spatial section of the wave front consists of two non-expanding 2-spheres of radius $b$ (given by $Z_2=\pm b$) spanned by the coordinates $(\rho,\phi)$. Their histories are depicted as two null lines in Fig.~\ref{fig_hyperboloid}, where the hyperboloid represents the AdS$_2$ factor of the LCBR background. Similarly, the trajectories $U=0$, $Z_2=\pm b$ are described by the solid lines in the conformal diagram of AdS$_2$ in Fig.~\ref{fig_bertotticonformal}. The function $H$ is singular at the poles $\rho=0,\pi b$, which are a remnant of the black hole and naked singularity of the original static spacetime. These describe null point particles sitting on the wave front (more comments below).

Metric~\eqref{final2} can also be  rewritten in terms of intrinsically four-dimensional null coordinates $(u,v)$ defined by (cf.., e.g., \cite{Ortaggio02})
\be
 U=\frac{u}{\Omega} , \qquad V=\frac{v}{\Omega}  , \qquad Z_2=(\pm)b\left(\frac{2}{\Omega}-1\right) , \qquad \Omega=1-\frac{1}{2}b^{-2}uv ,
 \label{uv}
\ee
leading to the line-element
\be
\d s^2=\frac{2\d u\d v+2H\delta(u)\d u^2}{\Omega^2}+\d\rho^2+b^2\sin^2\frac{\rho}{b}\d\phi^2 .
 \label{BRkruskal}
\ee
The function $H$ is still given by \eqref{final2}, where the first additive term, being a constant, can be removed by a discontinuous coordinate transformation (cf. eq.~(20) of \cite{Ortaggio02}). 
In these coordinates, the wave front sits at $u=0$ and  (timelike) infinity corresponds to $uv=2b^2$. The metric clearly admits the Killing vectors $\pa_v$ and $\pa_\phi$, and an additional smooth isometry is generated by $\pa_v-\frac{1}{2}a^{-2}u^2\pa_u$ (the latter exits thanks to the impulsive character of the metric \cite{Ortaggio02}). In the limit $b\to+\infty$, it reduces to the well-known Aichelburg-Sexl \pp wave propagating in Minkowski spacetime \cite{AicSex71} (upon noticing that the apparently diverging term $-2\sqrt{2}p\ln\frac{\rho^2}{4b^2}\delta(u)\d u^2$, which appears in the limit, can be regularized with a coordinate transformation of the form $v\mapsto v+k\theta(u)$, where $k$ is a constant).

Solution \eqref{BRkruskal} turns out to belong to a more general family of impulsive waves in the LCBR universe studied from a different viewpoint in \cite{Ortaggio02,OrtPod02}. It follows from \cite{Ortaggio02,OrtPod02} that the spacetime is of Petrov type N at $u=0$ (where the Weyl tensor is concentrated) and O elsewhere. The Ricci tensor is the same as in the LCBR background (a $\Phi_{11}$ component in Newman-Penrose notation), plus an additional distributional component due to the null point-particles localized at the poles of the spherical wave front. The latter is of the form \cite{Ortaggio02,OrtPod02} $\Phi_{22}\propto p\delta(u)\left[\delta(1-x)-\delta(1+x)\right]$, describing a pair of point particles with energy densities of opposite sign -- this reflects a similar situation with the two original static sources (cf. section~\ref{subsec_horizon} and \cite{AleGar96}). The fact that conical singularities disappear in the ultrarelativistic boost is related to the mass being rescaled to zero in the limit, so that $\a\to 1$ in \eqref{alpha} (cf. \cite{LouSan89a} for the boost of conical singularities in a different context).

It is also worth observing that spacetime \eqref{BRkruskal} belongs to the Kundt class \cite{Stephanibook,GriPodbook}, as can be seen by defining standard Kundt coordinates  
\be
 \tilde v=\frac{v}{\Omega} , \qquad \zeta=\sqrt{2}be^{i\phi}\tan\frac{\rho}{2b} , 
\ee
giving
\be
 \d s^2=2\d u\d \tilde v+\left[-b^{-2}\tilde v^2+2H\delta(u)\right]\d u^2+\frac{2\d\zeta\d\bar{\zeta}}{\left(1+\frac{1}{2}b^{-2}\zeta\bar{\zeta}\right)^2} , \qquad H=-\sqrt{2}p\frac{b^2}{b^2+l^2}\ln\frac{\zeta\bar\zeta}{2b^2} .
 \label{Kundt}
\ee
(Here we have removed the constant term originally appearing in $H$, as mentioned after eq.~\eqref{BRkruskal}.)
The Kundt vector field is defined by $\pa_{\tilde v}$ and is recurrent. Metric \eqref{Kundt} can thus be interpreted as the impulsive limit of exact non-expanding sandwich waves of finite duration, for which the component $2H\delta(u)\d u^2$ can be replaced by an arbitrarily profiled term $2H(u,\zeta,\bar\zeta)\d u^2$. A general family of purely {\em gravitational} waves (i.e., with $H$ of the form $H=f(u,\zeta)+\bar f(u,\bar\zeta)$ in \eqref{Kundt},  and $f(u,\zeta)$ arbitrary) was first obtained in \cite{GarAlvar84} (see also \cite{PodOrt03} for further comments and some extensions). For any choice of $H(u,\zeta,\bar\zeta)$, these spacetimes are in fact a subset of the ``degenerate'' Kundt spacetimes \cite{Coleyetal09} and all their scalar curvature invariants are constant (and coincide with those of the LCBR background) \cite{ColHerPel10}.

To conclude, it should be remarked that, prior to \cite{AleGar96}, a different solution describing the Schwarzschild black hole immersed in an external magnetic field was obtained in \cite{Ernst76a}. In that case, the background spacetime is given by the Bonnor--Melvin universe \cite{Bonnor54,Melvin64}. The ultrarelativistic boost of the Schwarzschild--Bonnor--Melvin black hole was studied in \cite{Ortaggio04}. More general gravitational waves in the Bonnor--Melvin spacetime were constructed earlier in \cite{GarMel92} and were also shown \cite{Ortaggio04} to belong to the Kundt class (but not to the subclass with constant curvature invariants).

\section*{Acknowledgments}

We thank George Alekseev for email correspondence. M.A. thanks the Centro de Estudios Cient\'{\i}ficos (CECS) for hospitality.
M.O. has been supported by research plan RVO: 67985840 and by the Albert Einstein Center for Gravitation and Astrophysics, Czech Science Foundation GA\v CR 14-37086G. The stay of M.O. at Instituto de Ciencias F\'{\i}sicas y Matem\'aticas, Universidad Austral de Chile has been supported by CONICYT PAI ATRACCI{\'O}N DE CAPITAL HUMANO AVANZADO DEL  EXTRANJERO Folio 80150028. M.A. is supported by Conicyt - PAI grant 79150061 and Fondecyt grant 11160945.

%
%

\end{document}